\documentclass[a4paper,12pt]{article}
\pdfoutput=1 
\usepackage{epsfig,graphicx,amsmath,amssymb}
\usepackage{mathrsfs}
\usepackage{chngcntr}
\counterwithin{figure}{subsection}
\counterwithin{table}{subsection}
\counterwithin{equation}{subsection}
\usepackage{url}
\usepackage{hyperref}
\newcommand{\ee}{e^{+} e^{-}}

\def \ee {e^+e^-}

\newcommand{\afb}{\mathcal{A}_{FB}}
\newcommand{\eetomumu}{e^{+}e^{-}\to\mu^{+}\mu^{-}}
\begin{document}

\thispagestyle{empty}

$\phantom{.}$

\hfill

\begin{center}
{\Large {\bf 15th Meeting of the Working Group on Rad. Corrections and MC Generators for Low Energies} \\
\vspace{0.75cm}}

\vspace{1cm}

{\large April 11, 2014 in Mainz, Germany}

\vspace{2cm}

{\it Editors:}
S.~E.~M\"uller (Dresden) and G.~Venanzoni (Frascati)

\vspace{2.5cm}

ABSTRACT

\end{center}

\vspace{0.3cm}

\noindent
The mini-proceedings of the 15th Meeting of the ``Working Group on Rad. Corrections and MC Generators for Low Energies'' held in Mainz on April 11, 2014, are presented. These meetings, started in 2006, have as aim to bring together experimentalists and  theorists working in the fields of meson transition form factors, hadronic contributions to $(g-2)_{\mu}$ and the effective fine structure constant, and development of MonteCarlo generators and Radiative Corrections for precision $\ee$ and $\tau$ physics.

\medskip\noindent
The web page of the meeting, which contains all talks, can be found at
\begin{center}
\url{https://agenda.infn.it/conferenceDisplay.py?confId=7800}
\end{center}

\vspace{0.5cm}


\newpage

{$\phantom{=}$}

\vspace{0.5cm}

\tableofcontents

\newpage

\section{Introduction to the $15^{th}$ Radio MontecarLow Working Group meeting}

\addtocontents{toc}{\hspace{1cm}{\sl H.~Czy\.z and G.~Venanzoni}\par}
\label{sec:Intro}
\vspace{5mm}

\noindent
H.~Czy\.z$^1$ and G.~Venanzoni$^2$

\vspace{5mm}

\noindent
$^1$ Institute of Physics, University of Silesia, 40007 Katowice, Poland\\
$^2$ Laboratori Nazionali di Frascati dell'INFN, 00044 Frascati, Italy\\
\vspace{3mm}

The importance of continuous and close collaboration between the experimental
and theoretical groups is crucial in the quest for
precision in hadronic physics.
This is the reason why the  
Working Group on ``Radiative Corrections and Monte Carlo Generators for Low Energies'' (Radio MonteCarLow)  was formed a few years ago bringing together experts (theorists and experimentalists) working in the field of low-energy $e^+e^−$ physics and partly also the $\tau$ community.
Its main motivation  was to understand the status and the precision of the Monte Carlo generators used to analyse the hadronic cross section measurements
obtained as well with energy scans as with radiative return, to determine luminosities, and whatever possible to perform tuned comparisons, {\it i.e.}
comparisons of MC generators with a common set of input parameters and experimental cuts. This  main effort was summarized in a report published in 2010~\cite{Actis:2010gg}.
During the years the WG structure has been enriched of more physics items 
and now it includes seven subgroups: Luminosity, R-measurement, ISR, 
Hadronic VP incl. $g-2$ and $\Delta \alpha$, gamma-gamma physics, FSR models, tau. 

During the workshop the latest achievements of each subgroups have been presented. 
A particular emphasis has been put to the recent evaluations of the Leading order and Light-by-Light hadronic contributions to the $g-2$ of the muon.
Finally the status of MC generators for R-measurement with energy scan, ISR, and tau decays has been discussed.
\medskip\noindent All the information on the WG can be found at the web page:
\begin{center}
\url{http://www.lnf.infn.it/wg/sighad/} 
\end{center}

\newpage

\section{Short summaries of the talks}

\subsection{Precision tests of unitarity in leptonic mixing}
\addtocontents{toc}{\hspace{2cm}{\sl J.~J.~van~der~Bij}\par}
\label{sec:vdBij}
\vspace{5mm}

J.~J.~van~der~Bij

\vspace{5mm}

\noindent
   Institut f\"ur Physik, Albert-Ludwigs Universit\"at Freiburg, Germany\\
\vspace{3mm}

First, the LHC has found no direct evidence for the existence of new physics beyond the standard model,
neither in the direct search nor in indirect effects in b-physics. As a consequence one should conclude that
major extensions of the standard model, for example supersymmetry, technicolor and the like, are ruled out,
or more politely are to be considered to be unlikely on experimental grounds. There is also an argument from
mathematical physics, that indicates that in the chiral sector at least the standard model is the only possible 
low energy theory\cite{vdbij1,vdbij2}. Therefore only the so-called minimalistic extensions of the standard model are possible.
These are extensions, that do not change the fundamental structure of the standard model in a major way, in particular
automatically not having flavor-changing neutral currents. These extensions basically consist of inert scalar multiplets,
that is multiplets not coupling to fermions, universal Z' bosons, coupling equally to all generations, and finally
sterile neutrinos. Such extensions are helpful in cosmology and can relatively easily explain a number of problems
like the existence of dark matter. Here we will be concerned with sterile neutrinos.

Second, the LHC has found the Higgs-boson of the standard model and its mass is now known to be about $ 126~{\rm GeV}$.
As a consequence predictions can be made within the standard model at the quantum level with an unprecedented
precision. Therefore we can now make tests on the model that were not possible before. In the past most tests 
assumed lepton universality implicitely, in order to be able to use the precision date from LEP to put limits
on the Higgs-boson mass. Now that the Higgs-boson mass is known, this is not sufficient anymore. One needs to combine
the LEP data with low energy measurements, that have also been improved. In the presence of sterile neutrinos,
the PMNS (Pontecorvo-Maki-Nakagawa-Sakata) matrix, that describes the mixing between the active neutrinos, coupled to
the weak interactions, is only part of the general neutrino mixing matrix. The full neutrino mixing matrix is of course 
unitary, but the PMNS matrix is only a 3x3 submatrix of the full matrix and therefore not unitary.
The deviations from unitarity can be described  by the parameters $\epsilon_e, \epsilon_{\mu}, \epsilon_{\tau}$,
that describe the amount of mixing of the $e, \mu, \tau$ neutrinos to the sterile neutrino sector. 
As a consequence the couplings of the fermions to the weak vector bosons are reduced by the $\epsilon$ parameters.

We performed a $\chi^2$ fit to the most precise data\cite{Basso}, consisting of $\tau$-decays, $\pi$-decays, test of the unitarity
of the Cabibbo-Kobayashi-Maskawa matrix, LEP-data and the W-boson mass. The combination of all data is sensitive to
values of ${\cal{O}} (10^{-3})$  of the $\epsilon$-parameters. We found a very good fit to the data with the following
characteristics. We found a 3-sigma deviation from zero 
$\epsilon_e = 2.5 \pm 0.8\, 10^{-3}$, with $\epsilon_{\mu}$ small and $\epsilon_{\tau}$ badly constrained.
In order to get an acceptable fit to the data, the measurement of the forward-backward asymmetry in the bottom quarks
at LEP, which has always been problematic, had to be left out. To come to a firm conclusion of course more experimental input
is required. From the experimental point of view the analysis is interesting, because results from practically every
high-energy accelerator contribute.  Improvements on measurements of the W-boson  mass, $\alpha(m_Z)$, $\tau$-decays and
the weak mixing angle will all contribute and can lead to an overall 5-sigma effect. Such experiments are ongoing
worldwide. The main point of this contribution is to point out, that these exeriments, though not designed for this
purpose, can contribute in a fundamental way to elucidate the question of the nature of dark matter in the universe.
Incidentally this proves, that the traditional differentiation between so-called discovery machines and so-called precision
machines is spurious. The motto here is :\\
\begin{center} PRECISION = DISCOVERY !.
\end{center}
\medskip
This work was supported  by the
Bundesministerium f\"ur Bildung und Forschung within the F\"orderschwerpunkt
\textit{Elementary Particle Physics}.

\newpage

\subsection{MC Generators for $e^+e^- \to$ hadrons at Low Energy}
\addtocontents{toc}{\hspace{2cm}{\sl S.~Eidelman, G.~Fedotovich, V.~Ivanov, A.~Korobov}\par}
\label{sec:Eidel}
\vspace{5mm}

S.~Eidelman, G.~Fedotovich, V.~Ivanov, A.~Korobov 

\vspace{5mm}

\noindent
      Budker Institute of Nuclear Physics SB RAS and  \\
Novosibirsk State University, Novosibirsk, Russia \\ 
\vspace{3mm}

 Two detectors, CMD-3 and SND, are now operated 
at the VEPP-2000 $e^+e^-$ collider with a goal of
high-precision measurements of various multihadronic 
cross sections~\cite{fedor}. Here we briefly describe several 
Monte Carlo (MC) generators created for these experiments. 

One of them considers the final state $K\bar{K}\pi$ with all three
charge combinations possible ($K^+K^-\pi^0,~K^0\bar{K}^0\pi^0,
~K^+\bar{K}^0\pi^-$). For each of them there are a few interfering
intermediate mechanisms. For example, for $K^+K^-\pi^0$ and
$K^0\bar{K}^0\pi^0$ there are at least three:  $\phi\pi^0$, $K^*(892)\bar{K}$ 
and direct three-body  $K\bar{K}\pi$ (production of higher mass $K^*$ 
states is also possible). It has been shown that interference 
effects can be rather large and should be carefully taken into account 
in the analysis of experiments on $e^+e^- \to  K\bar{K}\pi$~\cite{babarkkpi}
and $\tau^- \to (K\bar{K}\pi)^-\nu_\tau$~\cite{belletau,babartau}. 
     
For two-body leptonic and hadronic final states 
($e^+e^- \to e^+e^-,~\mu^+\mu^-$, \\
$\pi^+\pi^-,~K^+K^-$)~\cite{arbuz}
as well as for the two-photon annihilation 
($e^+e^- \to \gamma\gamma$)~\cite{gena} 
there is an MC generator MCGPJ providing a 0.2\% accuracy. 
This rather aggressive high accuracy is based on the
accuracy  claimed by the authors of the corresponding theoretical
evaluations and should be throughly checked by confronting experimental 
$\mu^+\mu^-\gamma$ and $\pi^+\pi^-\gamma$ events with a real photon
to results of MC simulation using the generator.    

We have also continued working on   a generic MC generator of
$e^+e^- \to$ hadrons below 2 GeV~\cite{fra13}. New processes were added,
matrix elements for more processes included. In view of the importance
of the energy range from 2 and 3 GeV for various applications like muon
anomalous magnetic moment, running fine structur constant etc. it is worth
discussing whether or not it is possible to extend the approach 
of this generator to the higher energy range. This task is not easy
because of a much higher number of final states accessible, but not
impossible after ISR measurements at Belle and BaBar as well as BESIII.




This work is supported by the Ministry of Education and Science of the
Russian Federation, the RFBR grants 12-02-01032, 13-02-00215   and 
the DFG grant HA 1457/9-1.

\newpage
\subsection{Towards a Precision Measurement of the Muon Pair Asymmetry in $\ee$ Annihilation at Belle and Belle\,II}
\addtocontents{toc}{\hspace{2cm}{\sl T.~Ferber}\par}
\label{sec:Ferber}

\vspace{5mm}

T.~Ferber

\vspace{5mm}

\noindent
Deutsches Elektronen--Synchrotron DESY, Hamburg, Germany\\
\vspace{5mm}

\begin{figure*}[!ht]
\includegraphics[width=14.0cm]{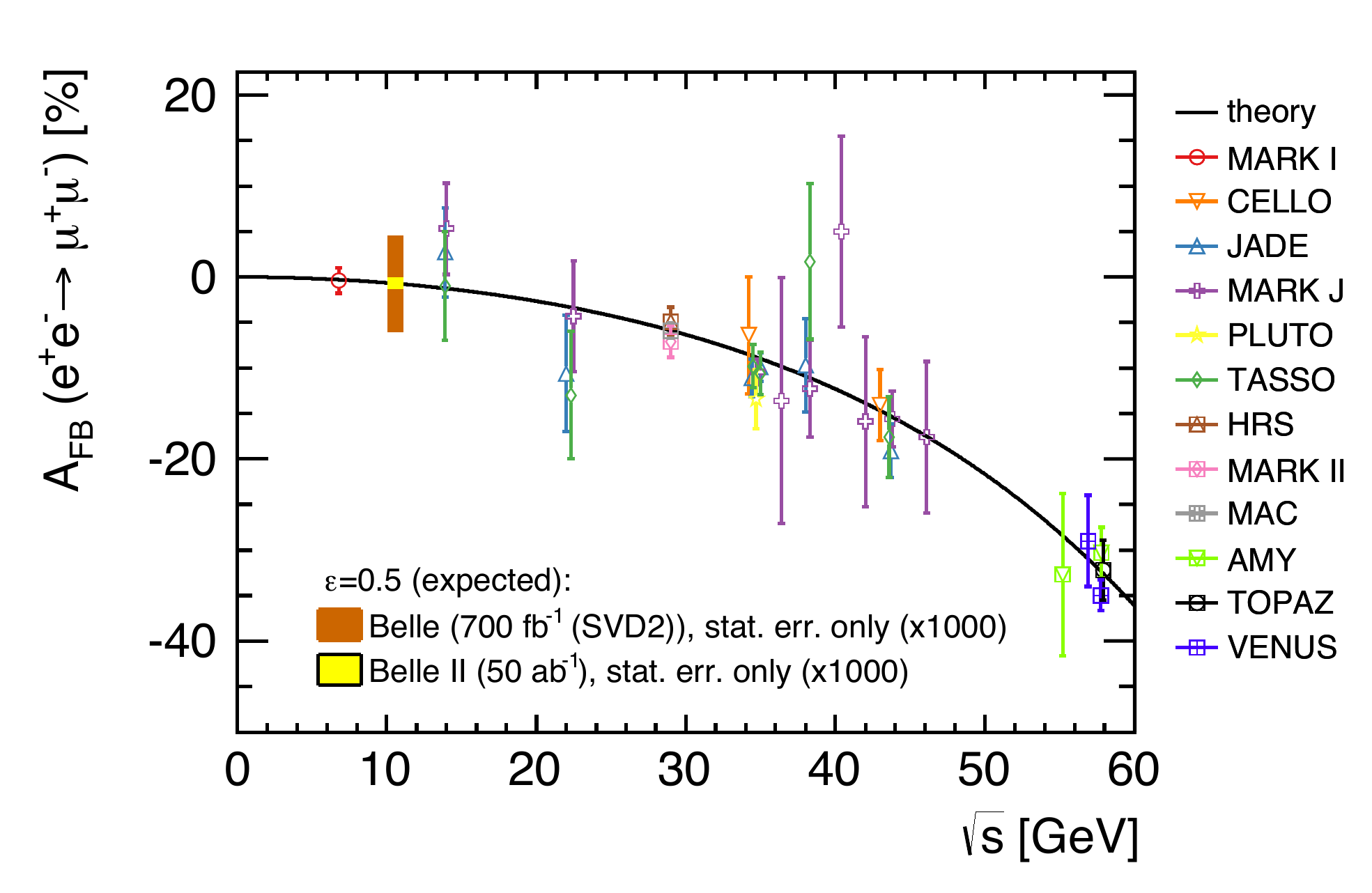}
\caption{\label{fig:afbmumu_history} Measurements of $\afb (\eetomumu)$ at different energies $\sqrt{s}$ corrected for QED effects by the respective authors (see \cite{venus} and references therein), theoretical SM prediction at lowest order and the expected Belle and Belle\,II statistical uncertainties (scaled up by a factor of 1000) at $\sqrt{s}=10.58$\,GeV.}
\end{figure*}

The process $\eetomumu$ is among the simplest reactions of the Standard Model (SM) where both quantum electrodynamics (QED) and electroweak (EW) predictions can be tested. The distribution of the polar angle $\theta^*$ of the outgoing leptons in the \hbox{center of mass} system, defined as the angle between the $e^+$ and the $\mu^{\pm}$, is expected to be asymmetric in the SM at Born level, caused by the interference of $\gamma$ and $Z$ exchange even at energies well below the Z pole, whereas lowest--order QED predicts a symmetric angular distribution. The forward--backward asymmetry is defined as
\begin{equation}
 \label{eq:afbdef}
 \afb^{\pm} \equiv \frac{N^{\pm}\left(\cos(\theta^*)\geq 0\right)-N^{\pm}\left(\cos(\theta^*)< 0\right)}{N^{\pm}\left(\cos(\theta^*)\geq 0\right)+N^{\pm}\left(\cos(\theta^*)< 0\right)},
\end{equation}
where $N^{\pm}(\cos(\theta^*))$ is the number of $\mu^{\pm}$ detected under the angle $\cos(\theta^*)$. At lowest order and neglecting initial and final state masses, the forward--backward asymmetry for $s\ll M_Z^2$ can be approximated as
\begin{equation}
 \label{eq:afbapprox}
 \afb^+(s)=-\afb^-(s)=\afb(s)\approx \frac{3 G_F}{4\sqrt{2}\pi\alpha}\frac{sM_Z^2}{s-M_Z^2}g_A^eg_A^{\mu},
\end{equation}
where $s$ is the squared center of mass energy, $G_F$ is the Fermi constant, $\alpha$ is the QED coupling constant, $M_z$ is the Z boson mass and $g_A^e$ and $g_A^{\mu}$ are the axial couplings of the electron and the muon. Previous measurements of $\afb$ are shown in Fig.~\ref{fig:afbmumu_history}. The forward--backward asymmetry $\afb$ is proportional to the $\rho$ parameter via $g_{A}^f=\sqrt{\rho_{f}}T_3^{f}$, where $T_3^{f}=1/2$ is the third component of the weak isospin and $f=e,\mu$. In the SM containing only Higgs doublets the $\rho$ parameter at lowest order is equal to unity \cite{bardin}. Deviation of the extracted $\rho$ parameter and its SM expectation after applying flavour--universal (u) and flavour--specific (f) virtual corrections
\begin{equation}
 \rho_{\text{f}}=1+\Delta\rho_{\text{u}}+\Delta\rho_{\text{f}}+\Delta\rho_{\text{new}}                                                                                                                                                                                                                                                                                                                                                                                                                                                                                                                                                                                                                                                                                                                                                                                                                                                                                                                                                                                                                                                                                                                                                                                                                                                                                                                                                                                                                                                                                        \end{equation}
can, e.g., be related to isospin violating New Physics through the oblique parameter $T$ \cite{peskin}, where the contribution to the low energy ($s\ll M_Z^2$) $\rho$ parameter is approximately given by \hbox{$\Delta\rho_{\text{new}} \approx \alpha_Z T$\,\cite{erler,erler2}} where $\alpha_Z\approx 1/128.945$ is the electromagnetic coupling at the Z pole. This measurement is unique in the sense that it probes axial--axial operators and allows an extraction of the oblique parameter $T$ that is independent of the oblique parameter $S$.\\

Using the existing unskimmed part of the Belle data with about $8\times10^8$ muon pairs (corresponding to about 0.7\,ab$^{-1}$), a precision measurement of $\afb((10.58\,\text{GeV})^2)$ with an expected statistical uncertainty of $\sigma(\afb)\approx10^{-4}$ (i.e. $\sigma(\afb)/\afb\approx1\,\%$) is possible, where the lowest--order SM prediction is $\afb^0((10.58\,\text{GeV})^2)\approx-0.77\,\%$. The Belle\,II experiment expects to collect about 50\,ab$^{-1}$ until 2023, thus reducing the expected statistical \hbox{uncertainty ($\sigma(\afb)\propto1/\sqrt{N}$)} to $\sigma(\afb)\approx10^{-5}$.\\

Apart from detector and background induced asymmetries and higher order weak virtual corrections, QED asymmetries of $\mathcal{O}(10^{-2})$ arise, mainly from \hbox{interference (IFI)} of initial and final state radiation and box diagrams. These asymmetries have to be corrected using Monte Carlo (MC) simulations. Ensuring both the theoretical uncertainty of $\afb$ and the implementation in the MC generators being understood at the level of $\sigma_{QED+EW}(\afb)\lesssim 10^{-5}$ make a detailed comparison of available (semi)analytical calculations and MC event generators mandatory. Such comparisons have been performed at energies around and above $M_Z$ for LEP1 and LEP2 energies before, but not at energies $s\ll M_Z^2$.\\

Two questions will be addressed using two independent (semi)analytical packages, one including the final state masses exactely (topfit) \cite{topfit1,topfit3} and ZFITTER \cite{zfitter1,zfitter2}, for idealized observables: Firstly, the final state muon mass beyond kinematic effects and secondly beyond one--loop electroweak corrections. If the effect of a non--zero final state muon mass is negligible for the required precision, ZFITTER can be used to study systematic effects after switching off beyond one--loop weak corrections in DIZET that includes approximations valid at the Z pole only and hence must be used carefully at $\sqrt{s}=10.58$\,GeV.\\

The ultimate step is to compare the two MC generators that are expected to principally provide the required precision needed to correct for QED effects and detector acceptance for a realistic event selection: KKMC\,4.19\footnote{Since KKMC uses the same DIZET library as ZFITTER, the same arguments apply concerning higher order weak corrections for $\sqrt{s}=10.58$\,GeV.} \cite{kkmc1} and PHOKHARA\,9.0 \cite{phokhara1}. For the QED parts relevant to the forward--backward asymmetry measurement, both generators differ mainly in the treatment of IFI, which is included in PHOKHARA\,9.0 up to complete NLO and in KKMC up to leading order plus coherent exclusive exponentiation \cite{kkmc2}. PHOKHARA lacks the inclusion of $Z$ exchange and $\gamma Z$ interference, whose implementation should be straightforward, though. Since for a MC statistical uncertainty below $\sigma(\afb)_{MC\,stat.}\approx10^{-5}$ one needs to generate $\mathcal{O}(10^{10})$ events, this is one of the reasons why it would be preferable to use tools like ZFITTER to evaluate systematics effects whenever applicable.\\

The measurement of $\afb$ with an absolute statistical uncertainty of $\sigma(\afb)\approx 10^{-5}$ at Belle\,II would allow precision tests of the SM, e.g. via the oblique parameter $T$, using the well defined forward--backward asymmetry below the Z pole if systematic uncertainties can be kept below $10^{-5}$. The required precision tag for QED corrections clearly is at the limit of currently available generators and needs to be understood. In addition, the planned work outlined above can serve as a benchmark for further generator studies for low energy $\ee$ colliders within the Radio MonteCarLow working group.

\newpage
\subsection{Direct production of $\chi_{c1}$ --- a $1^{++}$ charmonium at $e^+e^-$ machine}
\addtocontents{toc}{\hspace{2cm}{\sl Z.~Liu}\par}
\label{sec:Liu}
\vspace{5mm}

Zhiqing Liu
\vspace{5mm}

\noindent
Institute of Nuclear Physics, Johannes Gutenberg-University Mainz, Germany\\

\vspace{3mm}
Conventionally, an $e^+e^-$ annihilation machine only produce resonances with quantum numbers 
$J^{PC}=1^{--}$. This fact has been proved to be true since the discovery of the famous charmonium
state --- $J/\psi$~\cite{jpsi}. Now, several decades after, with the improving luminosity of $e^+e^-$ facilities, 
we are proposing to search for the direct production of a $1^{++}$ charmonium state --- $\chi_{c1}$, 
which sound impossible in $e^+e^-$ annihilation, but indeed can be accessible via two-photon exchange
process. We are aiming to give a $\Gamma_{ee}$ measurement for the $\chi_{c1}$ resonance, which
of course reflect the internal structure of this charmonium. Once the experimental approach has been
established, we can extend our study to another charmoniumlike state --- $X(3872)$~\cite{x3872}. 
The $X(3872)$ is a well-known charmoniumlike state with same quantum number as $\chi_{c1}$, 
however with an unknown nature. There are wide discussions whether this state is a hadron 
molecule~\cite{molecule} or tetraquark~\cite{tetraquark}. Our study of $X(3872)$ provides an
unique way to probe its internal structure, and finally will help reveal its true nature.

The BESIII experiment~\cite{bes3} located in Beijing is an advanced modern $e^+e^-$ machine.
The machine runs from $e^+e^-$ central-of-mass (cm) energy 2.0~GeV to 4.6~GeV, which covers
the full energy range of charmonium states. The designed luminosity is 
$1.0\times 10^{33} {\rm cm}^{-2} {\rm s}^{-1}$, which delivers high quality data roughly with 
average 15~pb$^{-1}$ per day. Our strategy is like this, the $\chi_{c1}$ resonance is produced directly
from $e^+e^-$ two-photon exchange process at peak, and subsequently decay into 
$\gamma J/\psi\to \gamma \mu^+ \mu^-$. The dominant backgrounds come from 
$\gamma_{ISR} J/\psi \to \gamma_{ISR} \mu^+ \mu^-$ and $\gamma_{ISR} \mu^+ \mu^-$ events, which
is estimated to be around 19~pb by MC simulation precisely~\cite{phokhara} and also confirmed by 
existing data sets at BESIII.
The $\Gamma_{ee}$ value of $\chi_{c1}$ resonance is estimated by VMD model to be 0.46~eV~\cite{vmd}.
Under this assumption, the bare signal cross section is estimated to be 637~pb. Considering 
initial-state-radiation (ISR) effect and beam energy spread~\cite{beam}, the real production cross section
reduced to 115~pb. Taking in the branching ratios and acceptance effect, the final effective cross section
is around 1.57~pb. With the signal to noise ratio 1.57/19=8.3\%, we expect to observe an evidence ($3\sigma$
significance) with 75~pb$^{-1}$ data, and a signal ($5\sigma$ significance) with 208~pb$^{-1}$ data.
Thanks to the good performance of BESIII, we are able to achieve $\chi_{c1}$ evidence with 5 days running,
and observation with 2 weeks running.

\newpage

\subsection{Monte Carlo Generators for the study of the process $e^+e^-\to 2(\pi^+\pi^-\pi^0)$ with the CMD-3 detector}
\addtocontents{toc}{\hspace{2cm}{\sl P.~A.~Lukin}\par}
\label{sec:Lukin}
\vspace{5mm}

P.~A.~Lukin

\vspace{5mm}

\noindent
Budker Institute of Nuclear Physics and Novosibirsk State University, Novosibirsk, Russia\\
\vspace{5mm}

The CMD-3 detector~\cite{Khazin:2008} at VEPP-2000 $e^+e^-$ collider~\cite{Danilov:1996} takes data in center-of-mass 
energy range $E_{cm} = 0.32$~--~$2.0$ GeV. During experimental seasons of 2011~--~2013 the luminosity integral of about 
60 $pb^{-1}$ has been collected. The analysis of the data is in progress and has been reported elsewhere (see, 
for example,~\cite{Ignatov:2014}). The study of the process $e^+e^-\to 3(\pi^+\pi^-)$ was finished and 
published~\cite{Solodov:2013}. Now the 6$\pi$ process in the channel $e^+e^-\to 2(\pi^+\pi^-\pi^0)$ is under study. 

\begin{figure}[h]
\begin{center}
\includegraphics[width=0.7\textwidth]{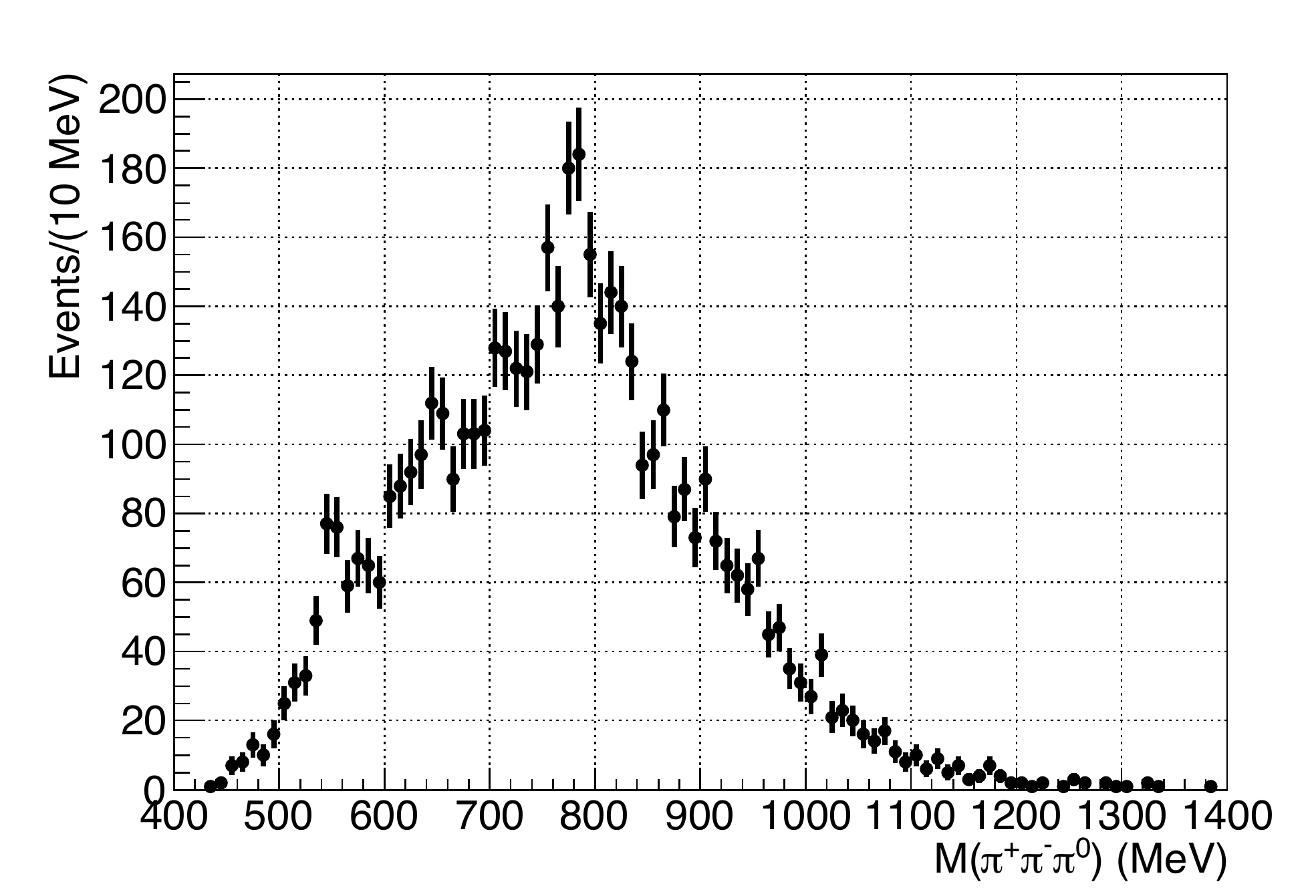}
\caption{Distribution of 2($\pi^+\pi^-\pi^0$) events over $\pi^+\pi^-\pi^0$ invariant mass at $E_{cm}$ = 1.72 GeV.}
\label{fig:fig1}
\end{center}
\end{figure}

In measurement of the cross section of any process it is important to describe correctly angular correlations between
particles in final state in order to correctly calculate registration efficiency of the process. So, it was our main 
goal in development of the Monte Carlo generators, although, studying of intermediate states is interesting itself.
We started to investigate distribution of events over invariant mass of three pions with zero charged. The corresponding 
plot is shown in Fig.~\ref{fig:fig1}. Clear signals from $\omega(782)$ and $\eta(545)$ mesons are seen. 
So, we need to take into account following intermediate contributions of $2(\pi^+\pi^-\pi^0)$ final state production:
\begin{eqnarray}
e^+e^- \to &\omega(782)\pi^+\pi^-\pi^0 & \to 2(\pi^+\pi^-\pi^0)\nonumber\\
e^+e^-\to  &\omega(782)\eta(545) &       \to 2(\pi^+\pi^-\pi^0)\nonumber.
\end{eqnarray}
As it was obtained from BaBar studying~\cite{Aubert:2006} and confirmed by the CMD-3 experiment~\cite{Solodov:2013} there 
is only one $\rho(770)$-meson production in 6$\pi$ final state. Therefore, in present work we also used Monte-Carlo 
generator $e^+e^-\to\rho(770)(4\pi)_{S-wave}$. Using these three processes we tried to describe mass and angular 
distributions of the $2(\pi^+\pi^-\pi^0)$ final state. 

\begin{figure}[h]
\begin{center}
\begin{tabular}{cc}
\includegraphics[width=0.48\textwidth]{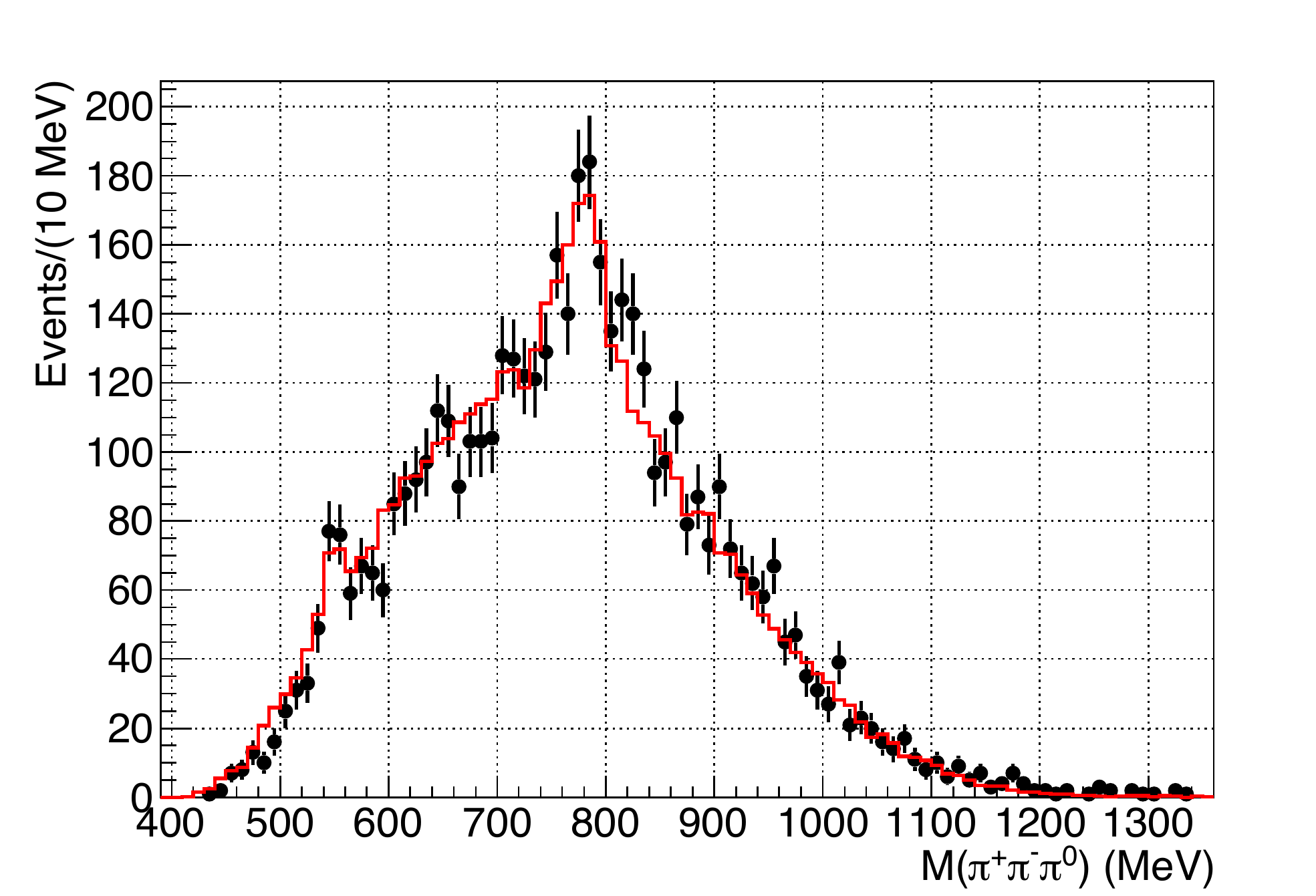} &
\includegraphics[width=0.48\textwidth]{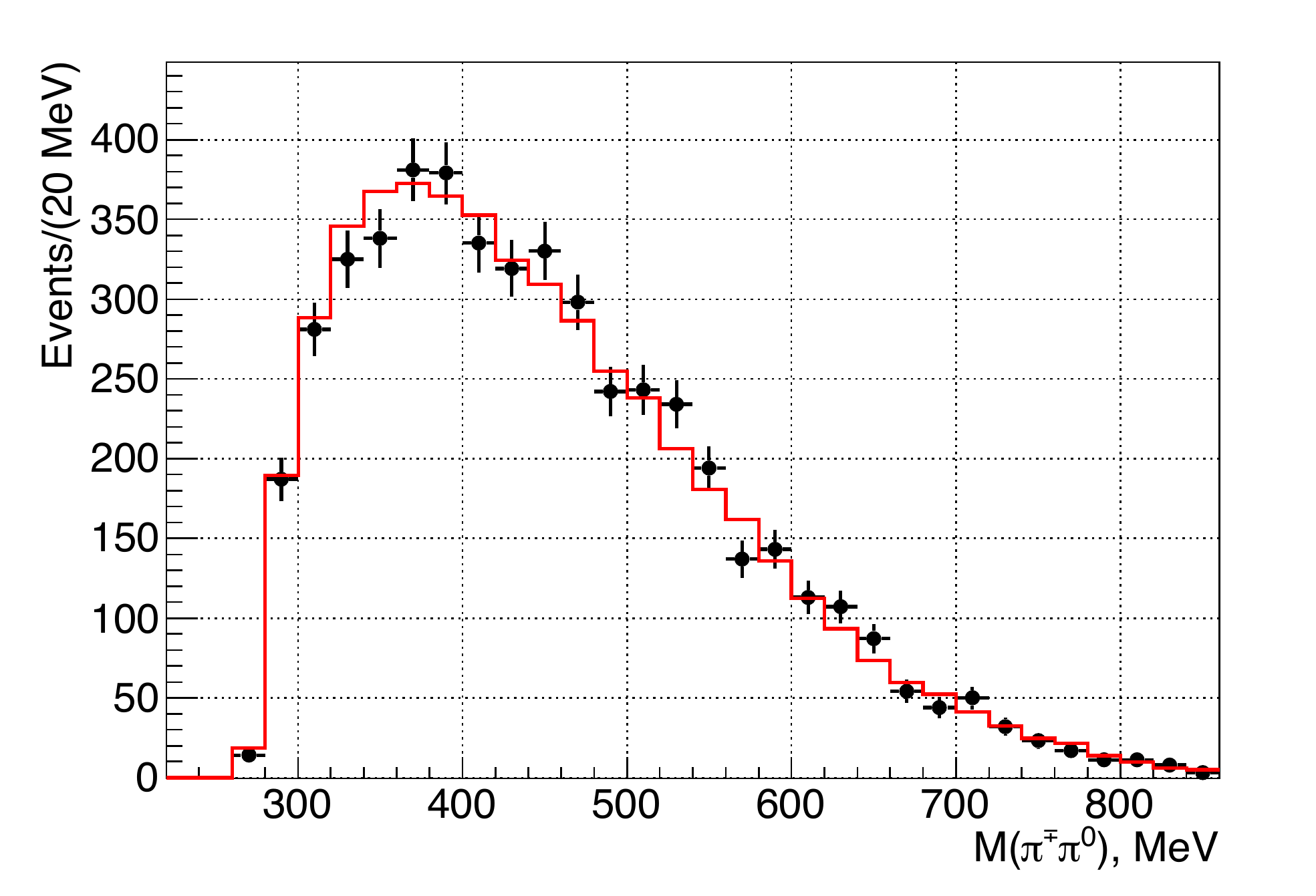} \\
\end{tabular}
\caption{On the left plot: Distribution of experimental data events over invariant mass of $\pi^+\pi^-\pi^0$ (points with 
errors) fitted with the sum of three contributions (histogram). On the right plot: Distribution 
of experimental data events over invariant mass of $\pi^{\pm}\pi^0$ (points with errors), described by the sum of three 
contributions (histogram). See details in the fit. }
\label{fig:fig2}
\end{center}
\end{figure}

\begin{figure}
\begin{center}
\includegraphics[width=0.7\textwidth]{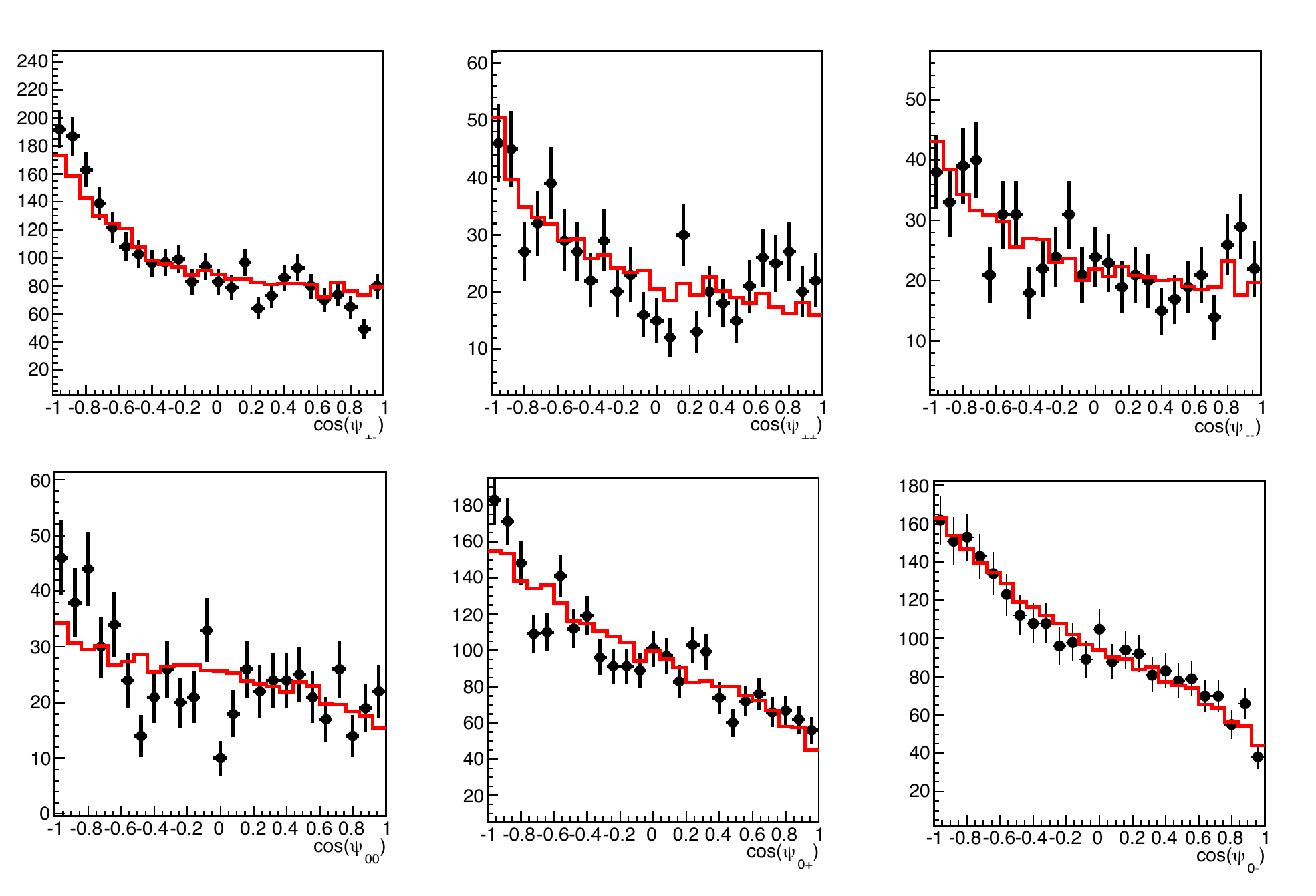}
\caption{Cosines of angles between (from left to right and from top to bottom) $\pi^+\pi^-$, $\pi^+\pi^+$, $\pi^-\pi^-$,
$\pi^0\pi^0$, $\pi^0\pi^+$ and $\pi^0\pi^-$ for experimental data (points with errors) and for simulation (histogram) of
three contributions --- $\omega(782)3\pi$, $\omega(782)\eta(545)$ and $\rho(770)(4\pi)_{S-wave}$.}
\label{fig:fig3}
\end{center}
\end{figure}

In Figure~\ref{fig:fig2}(left) one can see the distribution of experimental $2(\pi^+\pi^-\pi^0)$ events at $E_{cm} = 1.72$ 
GeV over invariant mass of $\pi^+\pi^-\pi^0$ (points with errors), fitted with the sum of simulated distributions of the
processes $\omega(782)\pi^+\pi^-\pi^0$, $\omega(782)\eta(545)$ and $\rho(770)(4\pi)_{S-wave}$. Fit result is shown as a 
histogram on the Fig.~\ref{fig:fig2}(left). The following fractions of the different contributions have been obtained:
\begin{eqnarray}
f_{\omega(782)3\pi}          & \sim & 60\% \nonumber\\
f_{\rho(770)(4\pi)_{S-Wave}} & \sim & 30\% \nonumber\\
f_{\omega(782)\eta(545)}     & \sim & 10\% \nonumber
\end{eqnarray}
The obtained values we applied to describe distribution of the experimental data events over invariant mass of 
$\pi^{\pm}\pi^0$ with the sum of three simulated process. The result is presented in Figure~\ref{fig:fig2}(right).
The fractions of different contributions, onbtained from the fit of the distribution in Figure~\ref{fig:fig2}(left) were
also used to describe angular correlations between particles in a 2($\pi^+\pi^-\pi^0$) final state. In Figure~\ref{fig:fig3} is shown
cosines of angles between (from left to right and from top to bottom) $\pi^+\pi^-$, $\pi^+\pi^+$, $\pi^-\pi^-$,
$\pi^0\pi^0$, $\pi^0\pi^+$ and $\pi^0\pi^-$ for experimental data (points with errors) and for simulation (histogram) of
three contributions --- $\omega(782)3\pi$, $\omega(782)\eta(545)$ and $\rho(770)(4\pi)_{S-wave}$ with relative fractions,
determined from the fit of distribution in Figure~\ref{fig:fig2}(left). Good agreement between experiment and simulation
can be seen at all plots in the Figure.

As result of the work presented here the mass and angular distributions of the 2$(\pi^+\pi^-\pi^0)$ events were described 
by the sum of three contributions $\rho(770)(4\pi)_{S-wave}$, $\omega(782)3\pi$ 
and $\omega(782)\eta(545)$  in center of 
mass energy range below 1.72 GeV. For energy above 1.72 new intermediate states should be investigated.

Author would like to thank organizers of XV RadioMonteCarlow meeting in April 2014 in Mainz (Germany) for support. 
This work is performed partly in the frame of program, described in~\cite{Actis:2010}.

\newpage

\subsection{Update on the combined estimate of the KLOE ISR measurements}
\addtocontents{toc}{\hspace{2cm}{\sl S.~E. M\"uller}\par}
\label{sec:Mueller}
\vspace{5mm}

S.~E.~M\"uller

\vspace{5mm}

\noindent
Institute of Radiation Physics, Helmholtz-Zentrum Dresden-Rossendorf, Germany\\

\vspace{3mm}
\setcounter{footnote}{0}
The KLOE experiment at the Frascati $\phi$-factory DA$\Phi$NE has published 4 data sets for the cross section of the process $e^+e^- \to \pi^+\pi^-$ below 1 GeV~\cite{Aloisio:2004bu,Ambrosino:2008aa,Ambrosino:2010bv,Babusci:2012rp}. As already described at the last meeting, work is in progress to combine the 3 datat sets  KLOE08~\cite{Ambrosino:2008aa}, KLOE10~\cite{Ambrosino:2010bv} and KLOE12~\cite{Babusci:2012rp}\footnote{The KLOE05~\cite{Aloisio:2004bu} data set is superseded by the more precise KLOE08~\cite{Ambrosino:2008aa} analysis} using the  method of the {\it best linear unbiased estimator} (BLUE)~\cite{Lyons:1988rp}. In ~\cite{Valassi:2003mu}, the method is extended to the combination of correlated measurements of several different observables, and an analytic solution is given to find the best estimates using the covariance matrix of the measurements. In our case,  this involves the construction of the covariance matrix for the 195 data points of the three KLOE measurements~\cite{Czyz:2013sga,Czyz:2013zga,KLOEurl}.  Since the presence of normalization errors in the $(195 \times 195)$ covariance matrix $\mathscr{M}_{ij}$  leads to a bias in the evaluation of the best estimates~\cite{dagostini}, the BLUE values are constructed using only the covariance matrix with statistical uncertainties. The covariance matrix that contains the normalization errors is then propagated properly a posteriori to the $(85 \times 85)$ covariance matrix of the best estimates.

As described in~\cite{Valassi:2003mu}, the BLUE method is equivalent to the problem of finding the estimates $\hat{x}_\alpha$ minimizing the sum
\begin{equation}
S = \sum_{\alpha=1}^N \sum_{\beta=1}^N \sum_{i=1}^{n}\sum_{j=1}^{n} \left[\mathscr{U}_{i\alpha}(y_i - \hat{x}_\alpha)\right]\mathscr{M}_{ij}^{-1}\left[\mathscr{U}_{j\beta}(y_j - \hat{x}_\beta)\right] \label{eq:chisq},
\end{equation}
in which the matrix $\mathscr{U}_{i\alpha}$ links the 195 data values from the KLOE publications $y_i$ to the 85 BLUE values $\hat{x}_\alpha$ (see~\cite{Valassi:2003mu,Czyz:2013sga}). $\mathscr{M}_{ij}$ is the (statistical) covariance matrix described above. Assuming that errors are Gaussian, the minimum of $S$ should be distributed as a $\chi^2$ with $(n-N)=195-85=110$ degrees of freedom. This can be used to estimate how consistent the individual measurements are with their combined estimates. In the present evaluation of the BLUE values, a value of $\chi^2_\mathrm{tot}/\mathrm{ndf} = 183/110$ is found, with a $\chi^2$-probability of $P\simeq 1.5 \times 10^{-5}$. This low value of probability can be justified by the fact that the method used to obtain the combined estimates only uses the statistical covariance matrix. Data points with large normalization uncertainties will therefore spoil the sum $S$ in eq.~\ref{eq:chisq}, therefore limiting the use of the $S$-value in as a consistency check for the BLUE method used to obtain the combined estimates. Keeping only the terms with $\alpha=\beta$, one can estimate the individual contributions $S_\alpha$ to eq.~\ref{eq:chisq} for each value of the $\hat{x}_\alpha$:
\begin{eqnarray}
S_\alpha & = & \sum_{i=1}^{n}\sum_{j=1}^{n} \left[\mathscr{U}_{i\alpha}(y_i - \hat{x}_\alpha)\right]\mathscr{M}_{ij}^{-1}\left[\mathscr{U}_{j\alpha}(y_j - \hat{x}_\alpha)\right]
\end{eqnarray}
Preliminary studies show that large values for $S_\alpha$ are found at the $\rho-\omega$ inteference region (where large uncertainties due to the procedure used to unfold the data from detector resolution are present) and around the value of 0.5 GeV$^2$ (where two points of the dominant KLOE08 data set pull the combined estimates away from the corresponding values of the KLOE10 and KLOE12 data). 

While the statistical contributions to the combined covariance matrix are 
under control, to conclude the work, a better understanding on the 
correlation between the systematic uncertainties of the KLOE08 and the 
KLOE12 analysis is needed. Currently, a full correlation between the two is 
assumed. It remains to be checked whether this assumption is valid.

\newpage

\subsection{Gradient method with re-weighted events and its implementation for {\tt TAUOLA} to fit the three pion mode}
\addtocontents{toc}{\hspace{2cm}{\sl J.~Zaremba}\par}
\label{sec:Zaremba}
\vspace{5mm}

J.~Zaremba

\vspace{5mm}
\noindent
Institute of Nuclear Physics, PAN, Krak\'ow, Poland\\
\vspace{5mm}

Recently, models based on the Resonance Chiral Lagrangian Theory~\cite{rchl} have been included into the Monte Carlo (MC) generator {\tt TAUOLA} for simulating hadronic $\tau$ decays. Models such as these must be tuned to the experimental data. For that purpose, a gradient method which uses MC re-weighting to morph the MC template used to tune the models has been adopted as one of the alternatives~\cite{web}. 

This method is motivated by its wide range of applications. For example, it can be used for data which has not been unfolded to account for the detector resolution, efficiency and experimental cuts. This is particularly useful for both multidimensional distributions and when fitting multiple channels at once.

In this approach, Monte Carlo sample is generated once and then re-used for fitting. When a set of the model parameters is changed, each event is given a weight which corresponds to the ratio of the matrix element calculated with the new set of parameters to the matrix element calculated at the time of generation. By using this re-weighting technique, one obtains both a numerically stable template which is suitable for fitting and reduces the computation time for generating the sample. 

Since the model used in {\tt TAUOLA} contains up to 15 parameters, taking into account time-consumption, scanning parameter-space randomly is not a reasonable option. In a first approximation, one can assume linear dependence on parameters. The assumption can always be made even if model is complicated. Through re-weighting we can morph our MC sample to any point of parameter space. Moreover, this technique allows us to construct any distribution available in the experiment. Using the Taylor expansion, a linear model can be constructed for a given point in the parameter space for each event. Since this simplified model holds for a linear combination of events, the simplified model can be constructed for each bin of the histograms. This enables standard tools like Minuit2 to be used to fit the simplified model to the experimental data. Then the procedure is repeated using this new set of parameters as starting point. As one can expect such a method is bound to circle around minima and requires further improvement.

As the most problematic issue is choosing the step size for the parameters such that one does not skip over the minima. In order to achieve this, one must incorporate information from the second derivatives. To address this problem, one must estimate the region of validity for the linear assumption of the simplified model. As a first estimate, this can be done by demanding that the ratio of the first to second order term in the Taylor expansion is much larger than a predefined value. The cross terms for the second order derivatives were neglected to reduce the computational time. This strategy provided a reasonable preliminary result in 10-20 iterations, a couple of days. However, full convergence can not be expected in this timescale. Due to the availability of the unfolded data from the experiments, further improvements to this method, such as an adaptive step-size, were not implemented.

In parallel with development of the discussed method, a semi-analytical fit was performed~\cite{Tauola} allowing cross-check with the approach in ~\cite{kukdm}. Even though full convergence, error evaluation, etc. is not yet possible for the Gradient method, results from both methods are very similar and they lead to same conclusions about the model used in {\tt TAUOLA}. Agreement represents technical cross-check of the methods. These studies will be continued to evaluate method performance. It will be important for future studies, when semi-analytical distributions and unfolded spectra of experimental data will not be available.

The author wishes to thank the organizers of the fifteenth meeting of the Radio Monte Carlo Working Group for support. Partially this project is financed from funds of Foundation of Polish Science grant POMOST/2013-7/12. POMOST Programme is co-financed from European Union, Regional Development Fund. This project is financed in part from funds of Polish National Science Centre under decisions  DEC-2011/03/B/ST2/00107. Part of the computations were supported in part by PL-Grid Infrastructure (http://plgrid.pl/) and were performed on ACK Cyfronet computing cluster (http://www.cyfronet.krakow.pl/).

\newpage

\section{List of participants}

\begin{flushleft}
\begin{itemize}

\item J.~J.~van~der~Bij, Albert-Ludwigs Universit\"at Freiburg, {\tt vdbij@physik.uni-freiburg.de }

\item S.~S.~Caiazza, Johannes Gutenberg-Universit\"at Mainz, {\tt caiazza@kph.uni-mainz.de}

\item C.~M.~Carloni-Calame, Pavia University {\tt carlo.carloni.calame@pv.infn.it}

\item H.~Czy\.z, University of Silesia, {\tt henryk.czyz@us.edu.pl }

\item A.~Dbeyssi, Johannes Gutenberg-Universit\"at Mainz, {\tt dbeyssi@kph.uni-mainz.de}

\item A.~Denig, Johannes Gutenberg-Universit\"at Mainz, {\tt denig@kph.uni-mainz.de}

\item S.~Eidelman, Novosibirsk State University, {\tt eidelman@mail.cern.ch }

\item T.~Ferber, DESY Hamburg, {\tt torben.ferber@desy.de}

\item Y.~Guo, Johannes Gutenberg-Universit\"at Mainz, {\tt guo@kph.uni-mainz.de}

\item A.~Hafner, Johannes Gutenberg-Universit\"at Mainz, {\tt hafner@kph.uni-mainz.de}

\item M.~Hoferichter, Albert Einstein Center for Fundamental Physics, Universit\"at Bern, {\tt hoferichter@itp.unibe.ch}

\item H.~Hu, IHEP Beijing, {\tt huhm@ihep.ac.cn}

\item G.~Huang, University of Science and Technology of China, {\tt hgs@ustc.edu.cn}

\item F.~Jegerlehner, Humboldt-Universit\"at zu Berlin, {\tt fjeger@physik.hu-berlin.de}

\item B.~Kloss, Johannes Gutenberg-Universit\"at Mainz, {\tt kloss@uni-mainz.de}

\item W.~Kluge, Institut f\"ur Experimentelle Kernphysik KIT, {\tt wolfgang.kluge@partner.kit.edu}

\item A.~Kupsc, Uppsala University, {\tt Andrzej.Kupsc@physics.uu.se }

\item Z.~Liu, Johannes Gutenberg-Universit\"at Mainz, {\tt liuz@uni-mainz.de}

\item P.~Lukin, Budker Institute of Nuclear Physics and Novosibirsk State University, {\tt P.A.Lukin@inp.nsk.su}
\item S.~E.~M\"uller, Helmholtz-Zentrum Dresden-Rossendorf, {\tt stefan.mueller@hzdr.de}
\item A.~Nyffeler, {\tt nyffeler@itp.unibe.ch}

\item C.~F.~Redmer, Johannes Gutenberg-Universit\"at Mainz, {\tt redmer@kph.uni-mainz.de}

\item M.~Ripka, Johannes Gutenberg-Universit\"at Mainz, {\tt ripka@uni-mainz.de}

\item B.~Shwartz, Budker Institute of Nuclear Physics, {\tt shwartz@inp.nsk.su}
\item E.~Solodov, Budker Institute of Nuclear Physics, {\tt E.P.Solodov@inp.nsk.su} 
\item T.~Teubner, University of Liverpool, {\tt thomas.teubner@liverpool.ac.uk }
\item M.~Unverzagt, Johannes Gutenberg-Universit\"at Mainz, {\tt unvemarc@kph.uni-mainz.de}

\item G.~Venanzoni, Laboratori Nazionali di Frascati dell'INFN, {\tt Graziano.Venanzoni@lnf.infn.it }

\item Y.~Wang, Johannes Gutenberg-Universit\"at Mainz, {\tt whyaqm@gmail.com}

\item S.~Wagner,  Johannes Gutenberg-Universit\"at Mainz, {\tt wagners@kph.uni-mainz.de} 

\item J.~Zaremba, Institute of Nuclear Physics, PAN, Krak\'ow, Poland, {\tt jakub.zaremba@ifj.edu.pl}

\end{itemize}
\end{flushleft}

\end{document}